\documentclass[useAMS]{mn2e}
\usepackage{lscape}
\usepackage{rotating}
\def\cm3{\hbox{cm$^{-3}$}}
\input psfig.sty
\input epsf.sty
\usepackage{graphicx}
\title[Evidence for the Strong Effect of Gas Removal on the Internal Dynamics of Young Stellar Clusters]
{Evidence for the Strong Effect of Gas Removal on the Internal Dynamics of Young Stellar Clusters}
\author[N.~Bastian \& S.P.~Goodwin] {N. Bastian$^1$ and S.P. Goodwin$^2$ \\
$^1$Department of Physics and
Astronomy, University College London, Gower Street, London, WC1E 6BT\\
$^2$Department of Physics and Astronomy, The University of Sheffield, Hicks Building, Hounsfield Road, Sheffield S3 7RH\\
}
\date{Accepted. Received; in original form}
\pagerange{\pageref{firstpage}--\pageref{lastpage}}
\pubyear{2005}
\begin{document}
\maketitle
\label{firstpage}
\begin{abstract}
%\vskip 70truemm
We present detailed luminosity profiles of the young massive 
clusters M82-F, NGC~1569-A, and NGC~1705-1 which show significant
departures from equilibrium (King and EFF) profiles.  We compare these 
profiles with those from $N$-body simulations of clusters which have 
undergone the rapid removal of a significant fraction of their mass due to gas
expulsion.  We show that the observations and simulations agree very
well with each other suggesting that these young clusters are
undergoing violent relaxation and are also losing a significant fraction of
their {\em stellar} mass.  

That these clusters are not in equilibrium can explain the
discrepant mass-to-light ratios observed in many young clusters with 
respect to simple stellar population models without resorting to
non-standard initial stellar mass functions as claimed for M82-F and 
NGC~1705-1. We also discuss the effect of rapid gas removal
on the complete disruption of a large fraction of young massive
clusters (``infant mortality'').  Finally we note that even bound 
clusters may lose $>$~50\% of their initial {\it stellar} mass due to 
rapid gas loss (``infant weight-loss'').

\end{abstract}
\section{Introduction}\label{intro}

It is thought that the vast majority of stars are formed in star
clusters (Lada \& Lada~2003).  During the collapse and fragmentation 
of a giant molecular cloud into a star cluster, only a modest 
percentage ($\sim 30 - 60$~\%) of the gas is turned into stars
(e.g. Lada \& Lada~2003).  Thus, during the initial phases of its lifetime,
a star cluster will be made up of a combination of gas and stars.
However, at the onset of stellar winds and after the first supernovae
explosions, enough energy is injected into the gas within the embedded
cluster to remove the gas on timescales shorter than a crossing time
(e.g. Hills~1980; Lada et al.~1984; Goodwin~1997a).  The 
resulting cluster, now devoid of gas, is far out of
equilibrium, due to the rapid change in gravitational potential energy
caused by the loss of a significant fraction of its mass.

While this process is fairly well understood theoretically
(e.g. Hills~1980; Mathieu~1983; Goodwin~1997a,b; Boily \& 
Kroupa 2003a,b), its effects have received little consideration in
observational studies of young massive star clusters.

In particular, many studies have recently attempted to constrain the
initial stellar mass function (IMF)\footnote{We note that estimates
  based on dynamical masses are in fact sensitive to the present mass
  function (MF).  However, it is assumed that for the young clusters
  studied the MF is likely to be a good representation of the {\it
    initial} mass function (IMF).} in clusters by studying the 
internal dynamics of
young clusters.  By measuring the velocity dispersion and half-mass
radius of a cluster, and assuming that the cluster is in Virial
equilibrium, an estimate of the dynamical mass can be made.  By then
comparing the ratio of dynamical mass to observed light of a cluster
to simple stellar population models (which require an input IMF)
one can constrain the slope or lower/upper mass cuts of the IMF
required to reproduce the observations.  Studies which have done such
analyses have found discrepant results, with some reporting
non-standard IMFs (e.g. Smith \& Gallagher~2001, Mengel et al.~2002)
and others reporting standard Kroupa~(2002) or Salpeter~(1955) type
IMFs (e.g. Maraston et al.~2004; Larsen \& Richtler~2004).

However, Bastian et al.~(2006) noted an age-dependence in how well
clusters fit standard IMFs, in the sense that all clusters $>$100~Myr
were well fit by Kroupa or Salpeter IMFs, while the youngest
clusters showed a significant scatter.  They suggest that this
is due to the youngest (tens of Myr) clusters being out of equilibrium,
hence undercutting the underlying assumption of Virial equilibrium
needed for such studies.

In order to test this scenario, in the present work we shall look at
the detailed luminosity profiles of three young massive clusters, namely
M82-F, NGC~1569-A, \& NGC~1705-1, all of which reside in nearby
starburst galaxies.  M82-F and NGC~1705-1 have been reported to have
non-standard stellar IMFs (Smith \& Gallagher~2001, McCrady et
al.~2005, Sternberg~1998).   Here we provide evidence that they are 
likely not in dynamical equilibrium due to rapid gas loss, thus 
calling into question claims of a varying stellar IMF.  NGC~1569-A
appears to have a standard IMF (Smith \& Gallagher~2001) based on
dynamical measurements, however we show that this cluster is likely
also out of equilibrium.  Throughout this
work we adopt ages of M82-F, NGC~1569-A, and NGC~1705 to be
 $60\pm20$~Myr (Gallagher \& Smith~1999), $12\pm8$~Myr (Anders et
al.~2004) and 10--20~Myr (Heckman \& Leitherer~1997) respectively.

Studies of star clusters in the Galaxy (e.g.~Lada \& Lada~2003)
as well as extragalactic clusters (Bastian et al.~2005a, Fall et
al.~2005) have shown the existence of a large population of
young ($<$~10-20~Myr) short-lived clusters.  The relative numbers 
of young and old clusters can only be reconciled if many young
clusters are destroyed in what has been dubbed ``infant-mortality''.
It has been suggested that rapid gas expulsion from young
cluster which leaves the cluster severely out of equilibrium would
cause such an effect (Bastian et al.~2005a).  We provide additional
evidence for this hypothesis in the present work.

The paper is structured in the following way.  In \S~\ref{data} and
\S~\ref{models} we present the observations (i.e. luminosity profiles)
and models of early cluster evolution, respectively.  In
\S~\ref{disc} we compare the observed profiles with our $N$-body 
simulations and in \S~\ref{conclusions} we discuss the implications
with respect to the dynamical state and the longevity of young clusters.

\section{Observations}\label{data}

For the present work, we concentrate on {\it F555W} (V) band
observations of M82-F, NGC~1569-A, and NGC~1705-1 taken with the {\it High-Resolution
Channel} (HRC) of the {\it Advanced Camera for Surveys} (ACS) on-board
the {\it Hubble Space Telescope} (HST).  The ACS-HRC has a plate scale
of 0.027 arcseconds per pixel.  All observations were taken from
the HST archive fully reduced by the standard automatic
pipeline (bias correction, flat-field, 
and dark subtracted) and drizzled (using the MultiDrizzle package -
Koekemoer et al.~2002) to correct for geometric distortions, remove
cosmic rays, and mask bad pixels.  The observations of M82-F are
presented in more detail in McCrady et al.~(2005).  Total exposures
were 400s, 130s, and  140s for M82-F, NGC~1569-A, and NGC~1705-1 respectively.

\subsection{Cluster profiles}

\begin{figure}%[htb]
\vspace{7.75cm}
\hspace{-2.cm}
\psfig{file=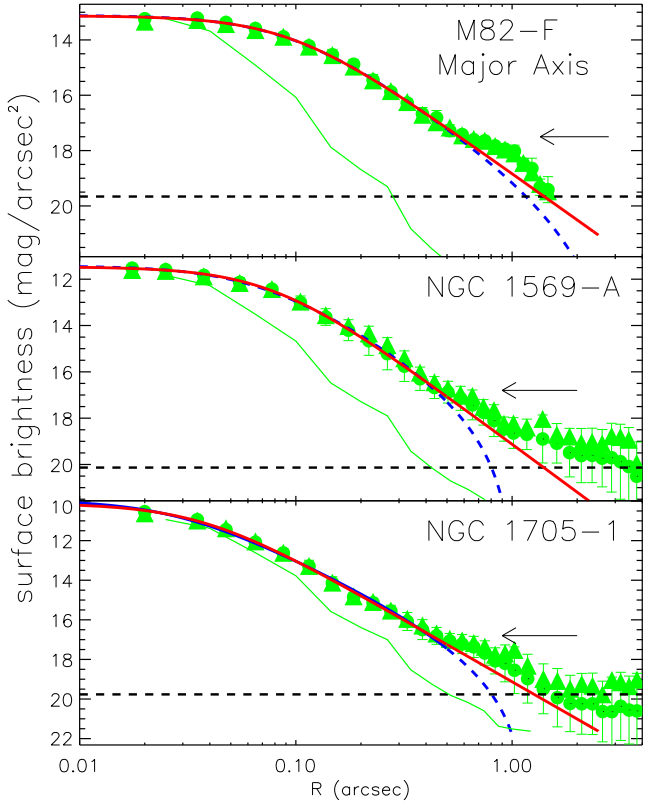,width=6.5cm,bbllx=100pt,bblly=267pt,bburx=440pt,bbury=530pt,angle=0}
\vspace{-2cm}
\caption{The luminosity profile of the young massive
  clusters M82-F, NGC~1569-A, NGC~1705-1.  The data are shown as
  filled green symbols, where the 
  circles are the mean of the flux at that radius and the
  triangles are the median of the flux.  Error bars represent
  the standard deviation of pixels in that annulus (based on the mean
  flux).  The PSF is shown 
  as a solid green line and the background value is the horizontal
  dashed line.  The best fitting King (blue/dashed) and EFF
  (red/solid) profiles
  are shown, see text.  The fits were carried out for $R <
  0.5''$. Note that all  clusters contain excess light at large
  radii relative to either of the profiles, the break from the EFF
  profile is noted by an arrow.}
\label{fig:obs}
\end{figure}

Due to the high signal-to-noise of the data, we were able to produce surface
brightness profiles for each of the three clusters on a per-pixel
basis.  The flux per pixel was background subtracted and transformed
to surface brightness.  The inherent benefit of using this technique, 
rather than circular apertures, is that it does not assume that
the cluster is circularly symmetric.  This is particularly important
for M82-F, which is highly elliptical (e.g. McCrady et al.~2005).  For
M82-F we took a cut through the major axis of the cluster.  The
results are shown in the top panel of Fig.~\ref{fig:obs}.  We note
that a cut along the minor-axis of this 
cluster as well as using different filters (U, B, and I - also from {\it
  HST-ACS/HRC} imaging) would not
change the conclusions presented in \S~\ref{disc}~\&~\S~\ref{conclusions}.

For NGC~1569-A and NGC~1705-1 we were able to assume circular symmetry 
(after checking the validity of this assumption) and hence we binned
the data as a function of radius from the centre.  The results for
these clusters are shown in the centre and bottom panels of
Fig.~\ref{fig:obs}, where 
the circular data points represent mean binning in flux and the
triangles represent median binning.  The standard deviation of the
binned (mean) data points is shown.   We also note that our conclusions
would remain unchanged (\S~\ref{disc}~\&~\S~\ref{conclusions}) if we
used the {\it F814W}~(I) {\it HST-ACS/HRC} observations.

We did not correct the surface brightness profiles for the PSF as the
effects that we are interested in happen far from the centre of the
clusters and therefore should not be influenced by the PSF.  In all
panels of Fig.~\ref{fig:obs} we show the PSF as a solid green line
(taken from an {\it ACS-HRC} observation of a star in a non-crowded
region).  The background of the area surrounding each cluster is shown
by a horizontal dashed line.

In order to quantify our results, we fit two analytical profiles to
the observed LPs.  The first is a King~(1962) function, which fits
well the Galactic globular clusters and is characterised by centrally
concentrated profiles with distinct tidal cut-offs in their outer
regions.  The second analytical profile used is an Elson, Fall, \&
Freeman~(EFF - 1987) profile, which is also centrally concentrated
with a non-truncated power-law envelope.  The EFF profile has been 
shown to fit young clusters in
the LMC better (EFF) as well as young massive clusters in galaxies
outside the local group (e.g. Larsen~2004; Schweizer~2004).  The best
fitting King and EFF profiles are shown as blue/dashed and red/solid lines
respectively.  The fits were carried out on all points within 0.5'' of
the centre of the clusters, i.e. the point at which, from visual
inspection, the profile deviates from a smoothly decreasing function.

As is evident in Fig.~\ref{fig:obs} all cluster profiles are well fit
by both King and EFF profiles in their inner regions.  {\it However,
none of the clusters appear tidally truncated, in fact all three clusters
display an excess of light at large radii with respect to the best
fitting power-law profile}.  The points of deviation from the best
fitting EFF profiles are marked with arrows.  This result will be
further discussed in \S~\ref{disc}. 

Due to the rather large distance of the galaxies as well as the
non-uniform background around the clusters presented here,
background subtraction is non-trivial.  However, we have checked the
effect of selecting different regions surrounding the clusters and
note that our conclusions remain unchanged.  We also note that in
the LMC, where the background can be much more reliably determined,
many clusters show excess light at large radii  (e.g.~EFF; Elson~1991; \&
Mackey \& Gilmore~2003).

\section{Simulations}\label{models}

We model star clusters using $N$-body simulations.  Star
clusters are constructed as Plummer (1911) spheres using the 
prescription of Aarseth et al. (1974) which require the Plummer
radius $R_P$ and total mass $M_P$ to be specified.  
Clusters initially contain 30000 equal-mass stars\footnote{Tests show
that the results do not depend on $N$, or the gravitational
softening.  This is because the dynamics we model are those of
violent relaxation to a new potential and so 2-body encounters are
unimportant.}.  Simulations were conducted on a GRAPE-5A special
purpose computer at the University of Cardiff using a basic
$N$-body integrator (the speed of the GRAPE hardware means that
sophisticated codes are not required for a simple problem such as this).

The expulsion of residual gas from star clusters has been modelled by
several authors (see in particular Lada et al. 1984; Goodwin 1997a,b;
Geyer \& Burkert 2001; Kroupa et al. 2001; Boily \& Kroupa 2003a,b).
The typical method is to represent the gas as an external potential
which is removed on a certain timescale.   Gas removal is expected to
be effectively instantaneous, i.e. to occur in less than a crossing time
(e.g. Goodwin 1997a; Melioli \& de Gouveia dal Pino 2006).  As 
such we require no gas potential, and can model the cluster as a 
system that is initially out of Virial equilibrium (equivalent to 
starting the simulations at the end of the gas expulsion).  The 
subsequent evolution is the violent relaxation (Lynden-Bell 1967) of
the cluster as it attempts to return to Virial equilibrium.

We define an {\em effective} star formation efficiency $\epsilon$
which parameterises how far out of Virial equilibrium the cluster is
after gas expulsion.  A cluster which initially contains $50$~\% stars 
and $50$~\% gas (i.e. a $50$~\% star formation efficiency) which is
initially in Virial 
equilibrium will have a stellar velocity dispersion that is a factor
of $\sqrt{2}$ - more generally $\sqrt{1/\epsilon}$ - too large to be
virialised after the gas is (instantaneously) lost.  We define the
efficiency as effective, as it assumes that the gas and stars are
initially in Virial equilibrium, which may not be true.

We choose as initial conditions, $R_P=3.5$~pc (corresponding to 
a half mass radius of $\sim 4$~pc) and $M_P/\epsilon = 5 \times 10^4$ 
or $7.5 \times 10^4 M_\odot$ (i.e. the total initial stellar plus gas
mass was $5 \times 10^4$ or $7.5 \times 10^4 M_\odot$) as 
representative of young massive star clusters.  In order to compare
the simulations with our observations we place the simulations at our
assumed distance of M82, namely 3.6~Mpc (assuming that it is at the
same distance as M81 - Freedman et al.~1994).

\section{A comparison of simulations and observations}\label{disc}

Previous simulations have shown that for $\epsilon<0.3$ clusters are totally
destroyed by gas expulsion, but for higher $\epsilon$ significant
(stellar) mass loss occurs, but a bound core remains (Goodwin 1997a,b;
Boily \& Kroupa 2003a,b).  For $\epsilon=0.4,0.5$ and $0.6$
respectively, $\sim 65, 35$ and $15$~\% of the initial stellar mass is
lost within $\sim 30$~Myr.  We confirm those results.

The escaping stars are not lost instantaneously, however.  Stars
escape with a velocity of order of the initial velocity dispersion of the
cluster, typically a few~km~s$^{-1}$.  Therefore, escaping stars will still be
physically associated with the cluster for $10$ -- $40$~Myr {\em after} gas
expulsion.  These stars produce a `tail' in the surface brightness
profile and produce the observed excess light at large radii.

We assume a constant mass-to-light ratio for the simulation and
convert the projected mass density into a luminosity and hence surface
brightness profile.  The normalisation of the surface brightness is
arbitrary and scaled so that the central surface brightness is similar
to that of the observed clusters.

\begin{figure}
\vspace{6.cm}
\hspace{-1.2cm}
\psfig{file=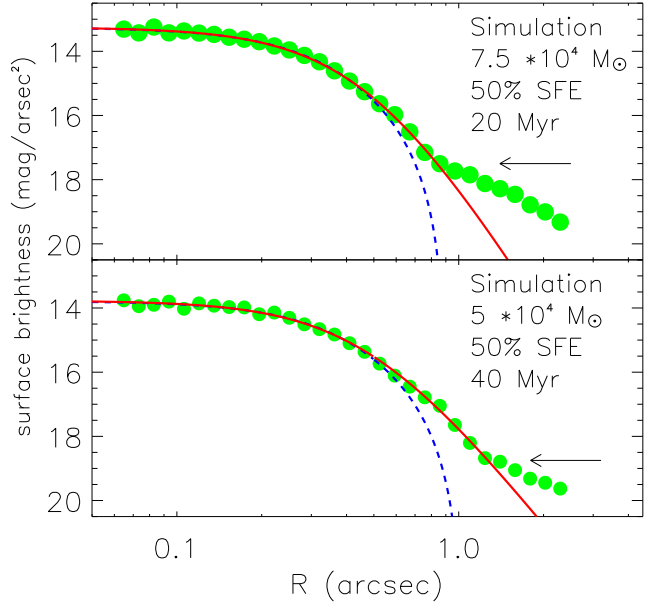,width=5cm,bbllx=100pt,bblly=267pt,bburx=440pt,bbury=530pt,angle=0}
\vspace{-1.5cm}
\caption{Two simulations of the effect of rapid gas loss on a young
  star cluster.  The green filled circles are the
  simulations, with a vertical scaling to match the observed
  clusters (see text).   We have transformed the simulations from
  parsecs to arcseconds assuming a distance to the clusters of
  3.6~Mpc.  The details 
  of each simulation are shown in the respective panels.  The blue/dashed
  and red/solid lines are the best fitting King and EFF profiles,
  respectively.  The fits were carried out for $R \leq 0.5$.  Note the
  large excess of the light at large radii
  with respect to the EFF profile.  The point of departure of the
  surface brightness profile from the EFF
  profile is noted by an arrow.}
\label{fig:model}
\end{figure}

Two of the simulations are shown in Fig.~\ref{fig:model}.  The filled
circles are the surface brightness of the simulated cluster, with the specific
parameters (total initial mass, $\epsilon$, and time since gas
expulsion) of the simulations shown.  We follow the same fitting technique as
with the observations, namely fitting King and EFF profiles
(dashed blue and solid red lines respectively) to the profile.  As was seen in the
observations, the simulations display excess light at large radii.

The detailed correspondence between the observations and simulations
presented here lead us to conclude that M82-F, NGC~1569-A and NGC~1705-1
display the signature of rapid gas removal and hence are {\em not in
dynamical equilibrium}.  In future works we will provide a large sample
of luminosity profiles of young massive extragalactic star clusters, as
well as a detailed set of models which can be used to constrain the
star formation efficiency of the clusters.  Here we simply note that
models with a SFE between 40-50\% best reproduce the observations.

Similar surface brightness profiles with an excess of light at large
radii are seen in young LMC clusters: see EFF and Elson~(1991)
in which many clusters clearly show these unusual profiles, and
also Mackey \& Gilmore~(2003) - in particular for R136.  These
profiles are also well matched by our simulations.  McLaughlin
  \& van der Marel~(2005) have compiled a data 
  base of structural parameters for young LMC/SMC clusters and compare
the M/L ratio from dynamical estimates to that predicted by simple
stellar population models (i.e. to check the dynamical state of the
young clusters).  However, the study was limited as the young clusters
tend to be of relatively low-mass, making it difficult to measure
accurate velocity dispersions.  Here we simply note that the five
clusters in their sample younger than 100~Myr all show significant
deviations in the M/L ratio, but also note that this may simply be due
to stochastic measurement errors.

\section{Implications and Conclusions}\label{conclusions}

It appears likely that the excess light at large radii seen in
many massive young star clusters is a signature of violent relaxation
after gas expulsion.  This suggests that these clusters have effective
star formation efficiencies of around $40$ -- $50$~\%, such 
that they show a significant effect, but do not destroy themselves rapidly.

\subsection{Virial equilibrium of young star clusters} 

\begin{figure}
\vspace{4.5cm}
\psfig{file=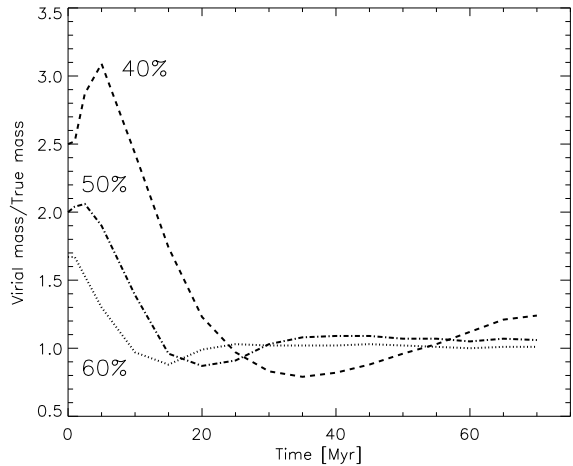,width=4cm,bbllx=100pt,bblly=267pt,bburx=440pt,bbury=530pt,angle=0}
\vspace{-1cm}

\caption{The variation of the virial-mass-to-true-mass ratio with
  time for effective star formation efficiencies of $40,~50$
  and~$60$~\%.  The virial mass is measured from the velocity
  dispersion of stars within 20~pc of the cluster centre and the true
  mass is the total mass of stars within 20~pc of the cluster.  Note
  that zero time is the time of gas expulsion, and so the cluster age
  is $\sim 5$~Myr greater than this time.}
\label{fig:virial}
\end{figure}

It should also be noted that the escaping stars are not just 
physically associated with a cluster in the surface brightness
profiles.  Measurements of the velocity dispersion of the cluster will
also include the escaping stars.  This will result in an artificially
high velocity dispersion that reflects the initial total stellar 
{\em and} gaseous mass.  Thus, mass estimates based on the 
assumption of stellar Virial equilibrium may be wrong by a factor of
up to three for 10--20~Myr after gas expulsion as is shown in
Fig.~\ref{fig:virial} for $\epsilon=40,~50$ and~$60$~\% clusters
(i.e.~at the ages of NGC~1569-A and NGC~1705-1).

Clusters with $\epsilon \sim 50$ --$60$~\% rapidly readjust to their
new potential and the virial mass estimates become fairly accurate
$10$ -- $15$~Myr after gas expulsion (i.e. for a cluster age of $15$ --
$20$~Myr).  However, for $\epsilon \sim 40$~\%, the virial mass
is significantly greater than the actual mass for $\sim 10$~Myr and
clusters do not settle into virial equilibrium for $\sim 50$~Myr
Indeed, between $30$ and $40$~Myr after gas expulsion the virial mass
estimate {\em underestimates} the total mass by up-to $30$~\% as 
the cluster has over-expanded.

A few recent studies have reported non-Kroupa~(2002) or non-Salpeter~(1955)
type initial stellar mass functions (IMF) in young star clusters (e.g. Smith
\& Gallagher~2001; Mengel et al.~2002).  These results were based on
comparing dynamical mass estimates (found by measuring the velocity
dispersion and half-mass radius of a cluster and assuming Virial
equilibrium) and the light observed from the cluster with simple
stellar population models (which assume an input stellar IMF).  Other
studies based on the same technique (e.g.~Larsen \& Ritchler~2004;
Maraston et al.~2004) have reported standard Kroupa- or Salpeter-type IMFs.

Recently, Bastian et al.~(2006) noted a strong age dependence on how
well young clusters fit SSP models with standard IMFs, with all
clusters older than $\sim100$~Myr being will fit by a Kroupa IMF.
Based on this age dependence, they suggested that the youngest star
clusters ($<80$~Myr) may not be in Virial equilibrium.  The observations
presented here strongly support this interpretation as M82-F and
NGC~1705-1 both seem to have been strongly affected by rapid gas
loss. While NGC~1569-A has been reported to have a Salpeter-type IMF
(Smith \& Gallagher 2001), the excess light at large radii suggests
that this cluster has also undergone a period of violent 
relaxation and stars lost during this are still associated with the
cluster even though its velocity dispersion correctly measures its mass..  

It should be noted that the obvious signature of violent relaxation in the
profile of M82-F suggests that it is at the lower end of its age
estimate of $60 \pm 20$~Myr (Gallagher \& Smith 1999), as 
by $40$ -- $50$~Myr the tail of stars
becomes disassociated from the cluster.  Another possibility is
that M82-F has been tidally shocked and has had a significant amount of
energy input into the cluster, thus mimicking the effects of gas
expulsion.  Whichever is the case, the tail of stars from M82-F -
whatever its age - is a signature of violent relaxation and strongly
suggests that it is out of virial equilibrium.

\subsection{Infant Mortality}

If a young star cluster has a low enough effective star formation
efficiency ($<30$~\%) it can become completely unbound and dissolve
over the course of a few tens of Myr.  This mechanism has been
invoked to explain the expanding OB associations in the Galaxy
(Hills~1980).  Recent studies of large extragalactic cluster
populations in M~51 (Bastian et al. 2005a) and NGC~4038/39 (Fall et
al.~2005) have shown a large excess of young ($<$10~Myr) clusters
relative to what would be expected for a continuous cluster formation
history.  Both of these studies suggest that the excess
of extremely young clusters is due to a population of short-lived
unbound clusters.  The rapid dissolution of these clusters has been
dubbed ``infant mortality''.

The observations and simulations presented here support such a
scenario.  If the star formation efficiency is less than 30\% - no
matter what the mass - the
rapid removal of gas completely disrupts a cluster (although see
Fellhauer \& Kroupa~2005 for a mechanism which can produce a bound
cluster with  $\epsilon \sim 20$\%).  Even if $\epsilon$ is large
enough to leave a bound cluster, the cluster may be out of equilibrium
enough for external effects to completely dissolve it, such as the
passage of giant molecular clouds (Gieles et al.~2006) or in the case
of large cluster complexes, other young star clusters.

Interestingly, gas expulsion often significantly lowers the {\em 
stellar} mass of the cluster even if a bound core remains (see 
\S~\ref{models}).  Thus, relating the observed mass function of 
clusters to the birth mass function needs to account not only for 
infant mortality, but also for `infant weight-loss' in which a cluster 
could lose $>50$~\% of its initial {\em stellar} mass in $<50$~Myr.

 The current simulations do not include either a stellar IMF, nor
the evolution of stars.  The inclusion of these effects do not
significantly effect the results  as the 
mass-loss due to stellar evolution is low compared to that due to 
gas expulsion (see Goodwin 1997a,b).  In particular, we do not expect the preferential loss of low-mass stars as these clusters are too young for equipartition to have occured, thus stars of all masses are expected to have similar velocities.  One caveat to this is the effect of primordial mass (hence velocity) segregation which may mean that the most massive stars are very unlikely to be lost as they have the lowest velocity dispersion.  We will consider such points in more detail in a future paper.

\section{Summary}

Observations of the surface brightness profiles of the massive young
clusters M82-F, NGC~1569-A, and NGC~1705-1 show a significant excess of light at
large radii compared to King or EFF profiles.  Simulations of the
effects of gas expulsion on massive young clusters produce exactly the
same excess due to stars escaping during a period of violent
relaxation.  Gas expulsion can also cause virial mass estimates to be
significantly wrong for several 10s~of~Myr.

These signatures are also seen in many other young star clusters
(e.g. Elson~1991; Mackey \& Gilmore~2003) and suggest that gas
expulsion is an important phase in the evolution of young clusters
that cannot be ignored.  In particular, this shows that claims of
unusual IMFs for young star clusters are probably in error as these
clusters are {\em not} in virial equilibrium as is assumed.

In future work we will further explore the dynamical state of
young clusters in order to constrain the star-formation efficiency
within the clusters.

\section*{Acknowledgments}

We would like to thank Mark Gieles and Francois Schweizer for
interesting and useful discussions, as well as Markus Kissler-Patig
and Linda Smith for critical readings of earlier drafts of the
manuscript.  The anonymous referee is thanked for useful suggestions
and comments.  This paper is based on observations with the NASA/ESA
{\it Hubble 
Space Telescope} which is operated by the Association of Universities
for Research in Astronomy, Inc. under NASA contract NAS5-26555.  SPG
is supported by a UK Astrophysical Fluids Facility (UKAFF)
Fellowship.  The GRAPE-5A used for the simulations was purchased on
PPARC grant PPA/G/S/1998/00642.

\bsp
\label{lastpage}
\end{document}